# Improvements in the contemporary implementation of photoemission spectroscopy

Swapnil Patil*

*Department of Physics, Indian Institute of Technology (Banaras Hindu University), Varanasi, 221005, India*

*Email: spatil.phy@itbhu.ac.in

## Abstract

In this short communication, we refined our previous manuscript to illustrate the corrections needed to the prevalent method for determining photoelectron counts in modern MCP-CCD-based ARPES detector systems and to implement our proposed 'new' photoelectron-detection method therein. We therefore provide a simple, practical solution for verifying changes in the ARPES spectrum using the 'new' method in modern ARPES spectrometers. The solution is simple enough to enable both the 'old' and 'new' methods within a single spectrometer, allowing a direct and rapid comparison of ARPES spectra collected under otherwise identical experimental conditions. We thus conclude our proposal.

## Background

In earlier versions of this article, we discussed ways to modify the ARPES technique to make it more sensitive to the photoelectron's many-body physics. Although we initially cited quantum phenomena—such as the photoelectron's wave function collapse [1], discussed, for example, in the famous Einstein, Podolsky, Rosen (EPR) paper [2]—to advocate for this modification, an equivalent but more practical motivation comes from a shortcoming in current ARPES electron detection (henceforth called 'old' ARPES). As an artefact of the prevalent detection process, old ARPES erroneously 'truncates' the photoelectrons' full many-body physics upon detection.

The main objective of the ARPES technique is to study the momentum-, energy-, and spin-dependent correlated electronic structure of materials [3]. Electron correlations arise from many-body physics, which typically 'renormalises' the properties of individual electrons. This renormalisation is integral to many-body phenomena and must be handled carefully when implementing ARPES to study them. In particular, the technique must distinguish photoelectrons that have been renormalised differently because they originate from different many-body states in the solid and record a distinct signal for each in the detector's final data. Only then can we distinguish photoelectrons with different renormalisations. The detector records a 'count value' for each photoelectron, along with its kinetic energy, emission angle and/or its spin. However, in 'old' ARPES, the count measured for each photoelectron is 'normalised' to a flat value of unity, regardless of the photoelectron's many-body origin or state of renormalisation. This normalisation defeats the very objective of the ARPES technique. Our proposal is therefore to remedy the situation by recording a 'renormalised' count value that can vary with the photoelectron's many-body origin. We call this approach 'new' ARPES. In this case, the renormalised count could be non-integral—for example, 1.2, 1.3, 0.9, or 0.7—instead of unity. This non-integral count value 'reflects' the many-body content of the photoelectron. For comparison, a 'bare' electron, with no many-body dressing, would naturally have an exact 'renormalised' count value of unity upon detection. A count value that differs from unity could therefore be considered a hallmark of many-body dressing accompanying the photoelectron. This implementation of ARPES is closer to its original objective.

A common argument against this idea is that photoelectrons ejected from a material lose their renormalisations as they travel through free space and are instead thought to behave like 'bare' electrons. This is now the prevailing view in the ARPES community. However, we strongly contest this picture. We argue that photoelectrons retain their many-body dressing as they travel through free space and therefore continue to exhibit renormalised properties, provided the measured signal is proportional to the renormalisation. The renormalised count value of a photoelectron is one such example. This claim, however, requires experimental verification using a 'new' ARPES spectrometer that records renormalised photoelectron counts.

**Discussion**

The renormalised count for a photoelectron can be determined in several ways. For example, within the 2D position-sensitive MCP-CCD detectors used in modern ARPES machines, it can be estimated by dividing the total charge accumulated in the CCD for a photoelectron hit at location $(x, y)$ by the total charge accumulated for a 'bare' electron hit at the same location. To determine the total accumulated charge for a photoelectron hit, the centroiding algorithm—which identifies the electron hit location from the CCD charge cloud—can be modified to also record the total charge, 'C', within the cloud. The accumulated charge for a 'bare' electron hit at the same location can be determined similarly, after calibrating the 2D detector with an electron gun (simulating a 'bare' electron), or by other methods. This calibration should be performed beforehand and would yield the accumulated 2D charge distribution $C^{cal}(x, y)$ for 'bare' electron hits across the detector before an ARPES experiment begins. In general, $C^{cal}(x, y)$ would differ for different pass energies, so it would need to be determined separately for each one. To estimate the renormalised photoelectron count, divide the total CCD charge from the photoelectron hit (C) by the $C^{cal}$ value at that hit's location $(x, y)$, i.e. $C/C^{cal}$, for the relevant pass energy. The resulting renormalised count, which is generally non-integral, should be used to increment the total photoelectron intensity at $(x, y)$ in the final ARPES data. Because renormalised count determination is purely a software change—requiring only a tweak to the centroiding algorithm and one additional mathematical operation—'old' and 'new' ARPES datasets could be recorded together within a single measurement. This would enable a clear comparison between the two types of ARPES and allow us to distinguish changes in the spectra that arise solely from the method used to determine the photoelectron count.

Lastly, some in the community may argue that the renormalisation of the electronic spectral function—widely thought to be visible in the 'old' ARPES spectra as well—results from many-body electronic correlations modulating the relative number of electron-emission events in particular energy and momentum regions. They may therefore believe that, despite ARPES measuring 'normalised' photoelectron counts, the spectra can still fully reveal the electrons' many-body physics. We take a different view. The theoretical many-body single-particle spectral function, $\boldsymbol{A}^{\pm}(\vec{\boldsymbol{k}}, \omega)$—with '+' denoting particle addition and '-' particle removal— is calculated using Green's function [4] and is believed to represent the electrons' many-body physics. When integrated over its full energy and momentum range, it should be constrained primarily by the number of photons that cause photoemission from the material. In that case, relative modulation of electron-emission events across energy and momentum regions may not, by itself, account for all the indications of the renormalised theoretical single-particle spectral function.

In either case, verifying and comparing the ARPES spectra by detecting photoelectrons at a renormalised count rate is purely a matter of scientific curiosity. Based on current knowledge, we anticipate substantial differences between the 'old' and 'new' ARPES spectra, particularly in materials with strong electron renormalisation effects. Rare-earth-based heavy-fermion/Kondo compounds, for example, exhibit some of the strongest electron-electron correlations reported in the literature. Their electronic renormalisation should therefore be greatest, making them the most suitable materials for observing this change.